\documentclass[reprint,aps,prc,amsmath,amssymb,nofootinbib,superscriptaddress]{revtex4-2}
\usepackage[colorlinks,linkcolor=blue,anchorcolor=blue,citecolor=blue]{hyperref}

\usepackage[utf8]{inputenc}
\usepackage[T1]{fontenc}

\usepackage{amsmath}
\usepackage{txfonts}
\usepackage{mathtools}
\usepackage{braket}
\usepackage{tensor}
\usepackage{xcolor}
\usepackage[abbreviations]{siunitx}
\usepackage{graphicx}
\usepackage{booktabs}

\usepackage{float}

\usepackage{hyperref}
\usepackage{cleveref}

\DeclareSIUnit{\fm}{\femto\meter}
\DeclareSIUnit{\MeVc}{\MeV\per\text{\ensuremath{c}}}

\newcommand{\tbd}[1]{{\color{black}#1}}

\begin{document}
 
 \title{The interplay of single-particle and collective motions in the low-lying states of $^{21}_\Lambda$Ne with quadrupole-octupole correlations}

\author{H. J. Xia}
\affiliation{Handan University, Handan 056005, P.R. China}

\author{X. Y. Wu} 
\email{xywu@jxnu.edu.cn}
\affiliation{College of Physics and Communication Electronics,
Jiangxi Normal University, Nanchang 330022, P.R. China}
\affiliation{School of Physics and Astronomy, Sun Yat-sen University, Zhuhai 519082, P.R. China} 

\author{H. Mei} 
\email{meihuayaoyugang@gmail.com}
\affiliation{Data Analysis and Research LLC, Okemos, Michigan 48864, USA}

\author{J. M. Yao} 
\email{yaojm8@sysu.edu.cn}
\affiliation{School of Physics and Astronomy, Sun Yat-sen University, Zhuhai 519082, P.R. China} 
 
\date{\today}

\begin{abstract}
 
The beyond mean-field approach for low-lying hypernuclear states  is extended by mixing the configurations associated with both single-particle and quadrupole-octupole collective excitations within the generator coordinate method based on a covariant density functional theory. The method is demonstrated in the application to the low-lying states of $^{21}_\Lambda$Ne, where the configurations with the $\Lambda$ hyperon occupying the first ($\Lambda_s$) and second ($\Lambda_p$) lowest-energy states are considered. The results indicate that the positive-parity states are dominated by the $\alpha+^{12}$C+$\alpha+\Lambda_s$ structure. In contrast, the low-lying negative-parity states are dominated by a strong admixture of $\alpha+^{16}$O+$\Lambda_s$ structure and $\alpha+^{12}$C+$\alpha+\Lambda_p$ structure due to the inclusion of octupole correlations.  As a result, the low-lying negative-parity states become much lower than what is expected from the previous studies without the mixing, and the electric multipole transition strengths are significantly quenched.  
\end{abstract}

\maketitle
 
 \section{Introduction}
 
  The hypernucleus is a self-bound quantum many-body system with one or more hyperons immersed in an ordinary nucleus composed of neutrons and protons. In a single $\Lambda$ hypernucleus, the $\Lambda$ hyperon can occupy any energy-allowed state as an impurity and thus provides a unique probe of the properties of atomic nucleus. Meanwhile, due to the short lifetime ($\sim10^{-10}$ s) of $\Lambda$ hyperon, it is difficult to perform $\Lambda$-nucleon ($\Lambda N$) scattering experiments~\cite{Hashimoto:2006,Gal:2016}. Thus, the $\Lambda N$ interaction is still poorly constrained, even though there is a proposal to collect more data at $e^+ e^-$ colliders~\cite{Dai:2022}, and remarkable progress has been achieved in the studies with lattice QCD~\cite{Beane:2007} and chiral effective field theory~\cite{Beane:2005,Haidenbauer:2013,Li:2016,Li:2018CPC,Song:2018,Song:2022}. Nevertheless, the lifetime of $\Lambda$ hyperons is much longer than the typical timescale of the strong interactions ($<10^{-22}$ s) and the typical half-lives ($\sim10^{-12}$ s) of nuclear excited states decaying via the electromagnetic interaction, one can produce $\Lambda$-hypernuclei at different energy states and measure their properties experimentally.  Therefore, the structure of $\Lambda$-hypernuclei  provides an important way to extract information on the $\Lambda N$ interaction in hypernuclear matter, the knowledge of which is relevant to solve the so-called {\em hyperon puzzle} in neutron stars~\cite{Lonardoni:2015PRL,Gandolfi:2015,Hagino:2016,Tolos:2020,Rong:2021,Tu:2022,Sun:2022}, namely, the difficulty to reproduce the measured maximal masses of neutron stars with the presence of hyperons.  
  
   Hypernuclei have been and will be studied more extensively with induced reactions of meson and electron beams at Japan Proton Accelerator Research Complex (J-PARC)~\cite{Koike:2019}, the Thomas Jefferson National Accelerator Facility (JLab), the Mainz Microtron (MAMI), the Research Center for Electron Photon Science (ELPH), Facility for Antiproton and Ion Research (FAIR), and the High Intensity Heavy-ion Accelerator Facility (HIAF)~\cite{Saito:2021,Zhou:2020AAPPS}. With the development of high-resolution germanium detector arrays for hypernuclear $\gamma$-ray spectroscopy,  low-lying states in several $p$-shell~\cite{Tamura:2000,Hashimoto:2006} and $sd$-shell~\cite{Yang:2018} hypernuclei have already been measured precisely. The next-generation facility J-PARC has already been in operation, opening up a new opportunity to perform high-precision hypernuclear $\gamma$-ray spectroscopy studies, especially the electric and magnetic dipole transition strengths~\cite{Aoki:2021}. The hypernuclear spectroscopic data that have been accumulated or to be measured contain rich information on hyperon-nucleon interactions in nuclear medium, and the impurity effect of hyperon particle on the structure of atomic nuclei.  To extract this information from data,  various hypernuclear models have been developed, including the valence-space shell model~\cite{Gal:1971,Gal:1972,Gal:1978,Millener:1985}, cluster~\cite{Motoba:1983} and few body models~\cite{Hiyama:2009PPNP}, ab initio no-core shell model~\cite{Wirth:2016,Wirth:2017,Le:2020}, and the Monte Carlo technique of impurity lattice effective field theory~\cite{Frame:2020}. 
 
 In addition to the hypernuclear models mentioned above, self-consistent mean field approaches~\cite{Bender:2003RMP} and nuclear energy density functional (EDF) theories~\cite{Meng:2006PPNP} have been extensively applied to study the bulky properties of hypernuclei~\cite{Rayet:1976,Mares:1993,LuHF:2002,Zhou:2007,Win:2008,Lu:2011,Tanimura:2012PRC,Schulze:2014,Xue:2015,Sun:2017,Liu:2018,Chen:2021SC,Rong:2021,Ding:2022PRC,Xue:2022}. In these studies, $\Lambda$ hyperon binding energies at different orbits have been well reproduced in hypernuclei from light to heavy mass regions, even though most of which are near closed shells and thus preserve spherical symmetry. It implies that the $\Lambda$ hyperon in these hypernuclei has a well-defined shell structure. In recent years, beyond-mean-field approaches have been developed for hypernuclear spectroscopy based on different nuclear energy density functionals, including anti-symmetrized molecular dynamics for hypernuclei  (HyperAMD)~\cite{Isaka:2011,Isaka:2012},  microscopic particle-rotor model~\cite{Mei:2014,Mei:2015}, and generator coordinate method for hypernuclei (HyperGCM)~\cite{Mei:2016PRC,Cui:2017} based on a Skyrme energy density functional or covariant density functional theory~\cite{Meng:2016Book,Meng:2020Hs,Wang:2022CTP}. In the latter, the configuration mixing of only axially deformed states was considered for the low-lying states of $^{21}_\Lambda$Ne, which has been studied  with microscopic cluster model~\cite{Yamada:1984} and HyperAMD~\cite{Isaka:2011}.  It has been found that the $\Lambda$ hyperon in the lowest energy state (labeled as $\Lambda_s$) weakly couples with the ground-state rotational band of the core nucleus $^{20}$Ne, forming a similar rotational band with almost degenerate doublet states. The $E2$ transition strength for the $2^+ \to 0^+$ transition in $^{20}$Ne is reduced by a factor ranging from 5\%~\cite{Cui:2017} to 13\%~\cite{Mei:2016PRC} by the $\Lambda_s$ hyperon. 
 
 Recently, we have extended the above HyperGCM for hypernuclear low-lying states  with the mixing of different quadrupole-octupole deformed configurations whose parity is allowed to be violated~\cite{Xia:2019}.  The broken symmetries are recovered with projections of parity, particle number, and angular momentum. It has been found that the $\Lambda$ hyperon  disfavors the formation of reflection-asymmetric molecular-like $\alpha$+$^{16}$O structure in $^{20}$Ne. One may expect that the electric octupole transition strengths are quenched, even though they are not investigated yet. In the meantime, we have found that the negative-parity states with the configuration $\ket{^{20}{\rm Ne}(K^\pi=0^-)}\otimes \ket{\Lambda_s}$ are close in energy to those with the  configuration $\ket{^{20}{\rm Ne}(K^\pi=0^+)}\otimes \ket{\Lambda_p}$.  In this work, we extend this framework further by mixing the above two types of configurations, examining the interplay of hyperon single-particle excitation and hypernuclear collective motions. This effect on nuclear energy spectra and electric multipole transition strengths will be studied in detail. \tbd{We note that  the effect of mixing the configurations  of $\ket{I^-}\otimes \ket{\Lambda_p}$ and $\ket{I^+}\otimes \ket{\Lambda_s}$  was studied before with an extended shell model for $^{12}_\Lambda$C and $^{16}_\Lambda$O~\cite{Motoba:1983NPA}.}
 
 The paper is organized as follows.  In Sect.~\ref{Sec.II}, we present the formalism of the extended HyperGCM for the low-lying states of single-$\Lambda$ hypernuclei with quadrupole and octupole correlations. In Sect.~\ref{Sec.results}, we present the energy spectra and the electric quadrupole and octupole transition strengths from the mixing of the two types of configurations with the $\Lambda$ placed on each of the first two lowest states. The results are discussed in comparison to those of calculations with one of the two types of configurations. A summary is given finally in sect.~\ref{Sec.summary}.

 \section{The framework} 
 \label{Sec.II}
 
 \subsection{The covariant density functional theory for quadrupole-octupole deformed hypernuclei}
  
 In the covariant density functional theory for $\Lambda$ hypernuclei, the effective nucleon-nucleon ($NN$) and $\Lambda N$ interactions are described by the Lagrangian density of relativistic point-coupling model, where the $\Lambda N$ interaction part is given by~\cite{Tanimura:2012PRC},
 \begin{eqnarray} 
{\cal L}^{N \Lambda}&=& -\alpha_S^{(N\!\Lambda)}(\bar{\psi}^{N}\psi^N)(\bar{\psi}^{\Lambda}\psi^{\Lambda})\nonumber\\ 
&&
-\alpha_V^{(N\!\Lambda)}(\bar{\psi}^{N}\gamma_{\mu}\psi^N)
(\bar{\psi}^{\Lambda}\gamma^{\mu}\psi^{\Lambda})\nonumber\\ 
&&-\delta_S^{(N\!\Lambda)}(\partial_{\mu}\bar{\psi}^{N}\psi^N)
(\partial^{\mu}\bar{\psi}^{\Lambda}\psi^{\Lambda}) \nonumber\\
&& -\delta_V^{(N\!\Lambda)}(\partial_{\mu}\bar{\psi}^{N}\gamma_{\nu}\psi^N)
(\partial^{\mu}\bar{\psi}^{\Lambda}\gamma^{\nu}\psi^{\Lambda})\nonumber\\
 &&+
\alpha^{(N\!\Lambda)}_T(\bar{\psi}^{\Lambda}\sigma^{\mu\nu}\psi^{\Lambda})
(\partial_{\nu}\bar{\psi}^{N}\gamma_{\mu}\psi^N).
\end{eqnarray} 
 The $\psi^{N/\Lambda}$ represents nucleon and $\Lambda$ hyperon fields, respectively. Nucleons and $\Lambda$ hyperon are further approximated as independent particles trapped in each potential determined by the densities and currents in a self-consistent manner. All the fields are taking the expectation values of the mean-field state whose wave function $\ket{\Phi^{(N\Lambda)}_{n}(\mathbf{q})}$  can be factorized as a product of the nuclear part and hyperon part, 
\begin{equation}
\label{eq:hypernuclear_config}
    \ket{\Phi^{(N\Lambda)}_{n}(\mathbf{q})}
    = \ket{\Phi^{N}(\mathbf{q})} \otimes\ket{\varphi^{(\Lambda)}_n(\mathbf{q})}
\end{equation}
and is determined by the Ritz variational principle. To generate the wave function of a $\Lambda$-hypernucleus with the correct average number of particles and different multipole deformation parameters $\mathbf{q}$, we add a linear term on the particle-number operator $\hat N$, and quadratic constraint terms on the quadrupole and octupole moments
 \begin{eqnarray}
 \label{Dirac}
 \delta \bra{\Phi^{(N\Lambda)}_{n}(\mathbf{q})} \hat H 
 && - \sum_{\tau=n, p} \lambda_\tau \hat N_\tau \nonumber\\
 && - \sum_{\lambda=1, 2,3} C_\lambda (\hat Q_{\lambda0} - q_\lambda)^2  \ket{\Phi^{(N\Lambda)}_{n}(\mathbf{q})}=0.
\end{eqnarray}
 The Lagrange multiplier $\lambda_\tau$ is determined by the constraint $\langle q \vert \hat N_\tau\vert q\rangle=N(Z)$. The $C_\lambda$ is the stiffness parameter. The center-of-mass coordinate  is imposed by the constraint $\langle \Phi^{(N\Lambda)}_{n} \vert \hat Q_{10} \vert \Phi^{(N\Lambda)}_{n}\rangle=0$. The $\hat N_\tau$ and $\hat Q_{\lambda 0}\equiv r^\lambda Y_{\lambda 0}$ are particle number and mass multipole moment operators, respectively. The deformation parameters $\beta_{\lambda\mu}$ are defined as
\begin{equation}\label{deformation}
  \beta_{\lambda\mu} 
  =\dfrac{4\pi}{3A R^\lambda} \bra{\Phi^{(N\Lambda)}_{n}(\mathbf{q})} \hat Q_{\lambda\mu} \ket{\Phi^{(N\Lambda)}_{n}(\mathbf{q})},
\end{equation}
with $A$ representing the mass number of the nucleus,  $\quad R=1.2A^{1/3}$ fm.  For simplicity, the deformation parameters $\mathbf{q}=(\beta_{20}, \beta_{30})$ are considered. In this case, the hyperon wave function $\ket{\varphi^{(\Lambda)}_n(\mathbf{q})}$ can be characterized  with the quantum number $\Omega_\Lambda$ which is the component of the angular momentum of the hyperon along the $z$-axis. Besides, the hypernuclear system composed of an even-even nuclear core plus a single $\Lambda$ hyperon is considered. The symbol $n$ distinguishes different hyperon states. The $\Lambda$ is placed in one of the two lowest energy states and is labeled as $\Lambda_s$ and $\Lambda_p$, respectively, even though the $\Lambda$ wave function is generally an admixture of states with different orbital angular momenta due to nonzero quadrupole-octupole deformation. 

 In this work, we employ the relativistic point coupling energy functional  PC-F1~\cite{Burvenich:2002PRC} parameterization for the $NN$ effective interaction and the PCY-S2~\cite{Tanimura:2012PRC} for the $\Lambda N$ effective interaction.  We note that the calculation using PC-PK1~\cite{Zhao:2010PRC} for the $NN$ effective interaction does not change the topology of the energy surface for the nuclear core, even though the results could be somewhat different quantitatively. See the study in Ref.~\cite{Zhou:2016PLB}. 

 \subsection{The generator coordinate method for hypernuclei}
In the extended version of HyperGCM for single-$\Lambda$ hypernucleus, the wave function of a hypernuclear state with spin-parity $J^\pi$ is generally constructed as a superposition of quantum-number projected hypernuclear mean-field states,
\begin{eqnarray}
\label{GCM:wf}
\ket{\Psi^{J^\pi}_{\alpha} }
= \sum _{n,\mathbf{q},K} f^{JK\pi}_{\alpha}(\mathbf{q},n)\ket{NZ JK \pi; \mathbf{q},n},
\end{eqnarray}
where the symbol $\alpha$ labels the quantum numbers of the state other than the spin parity $J^\pi$.  The basis function $\ket{NZ JK\pi; \mathbf{q},n}$ for the many-body state $\ket{\Psi^{J^\pi}_{\alpha} }$ is determined as follows.
\begin{eqnarray}
\label{basis_function}
\ket{NZ JK\pi; \mathbf{q},n} =  \hat{P}^J_{MK}\hat{P}^N \hat{P}^Z\hat P^\pi \ket{\Phi^{(N\Lambda)}_{n}(\mathbf{q})},
\end{eqnarray}
where $\hat P^{J}_{MK}$,  $\hat{P}^{N, Z}$, and $\hat P^\pi$ are the projection operators that extract the component with the right angular momentum $J$, neutron number $N$, proton number $Z$, and parity $\pi$, 
\begin{subequations}
\begin{align}
    \hat P^{J}_{MK}&=\dfrac{2J+1}{8\pi^2}\int d\Omega D^{J\ast}_{MK}(\Omega) \hat R(\Omega),\\
    \hat P^{N_\tau} &= \dfrac{1}{2\pi}\int^{2\pi}_0 d\varphi_{\tau}  e^{i\phi_{\tau}(\hat N_\tau-N_\tau)},\\
  \hat P^\pi &= \dfrac{1}{2}(1+\pi\hat P),
\end{align} 
\end{subequations}
where the operator $\hat P^J_{MK}$  extracts from the intrinsic state $|\Phi^{(N\Lambda)}_{n}(\beta_{20},\beta_{30})\rangle$ the component whose angular momentum along the intrinsic axis $z$ is given by $K$. The $\hat P^{N_\tau}=\hat P^{N,Z}$ and $\hat P^\pi$ are the projection operators of particle numbers ($N, Z$) and parity $\pi=\pm1$, respectively. For the single-$\Lambda$ hypernucleus with an even-even nuclear core, the total angular momentum $J$ is a half-integer number. Besides, as discussed in Eq.(\ref{Dirac}), the symmetry of rotation invariance with respect to the $z$-axis is imposed in the intrinsic configurations $\ket{\Phi^{(N\Lambda)}_{n}(\mathbf{q})}$, there is no $K$ mixing. Therefore, the $K$ quantum number in Eq.(\ref{basis_function}) is identical to $\Omega_\Lambda$. For the mean-field states with the hyperon in a $\Omega_\Lambda$ configuration, the angular momentum $J$ of the projected state takes the value of $\vert \Omega_\Lambda\vert, \vert \Omega_\Lambda\vert+1, \cdots$.

The weight function $f^{JK\pi}_{\alpha}(\mathbf{q},n)$ in Eq.(\ref{GCM:wf}) is determined by the variational principle, which in the case of without $K$-mixing leads to the following generalized eigenvalue equation,
\begin{eqnarray}
\label{HWE}
\sum_{n',\mathbf{q}'}
\left[{\cal H}^{JK\pi}_{nn'}(\mathbf{q},\mathbf{q}') -E^{JK\pi}_{\alpha}
{\cal N}^{JK\pi}_{nn'}(\mathbf{q},\mathbf{q}')\right]
f^{JK\pi}_{\alpha}(\mathbf{q}',n')=0.
\end{eqnarray}

The norm kernel ${\cal N}$, Hamiltonian kernel ${\cal H}$, and the kernel of electric multipole operators can generally be written in the following form.
\begin{eqnarray}
\label{eq:kernels}
&&{\cal O}^{JK\pi}_{nn'}(\mathbf{q},\mathbf{q}') \nonumber\\
&=&
\bra{NZ JK\pi; \mathbf{q},n}\hat O \ket{NZ JK\pi; \mathbf{q}',n'} \nonumber\\
&=& \frac{2J+1}{8\pi^2}\int  D^{J\ast}_{KK} (\Omega)d\Omega
\int^{2\pi}_0 \frac{e^{-i\phi_N N}}{2\pi}d\phi_N
\int^{2\pi}_0 \frac{e^{-i\phi_ZZ}}{2\pi} d\phi_Z
\nonumber\\
&& 
\times\bra{\Phi^{(N\Lambda)}_{n}(\mathbf{q})} \hat O\hat R(\Omega)e^{i\phi_N\hat N}e^{i\phi_Z\hat Z}\left(\frac{1+\pi\hat P}{2}\right) \ket{\Phi^{(N\Lambda)}_{n'}(\mathbf{q}')}
\end{eqnarray}
with $\hat{O}=1, \hat{H}$, and $\hat{Q}^{(e)}_{\lambda \mu}\equiv er^\lambda Y_{\lambda\mu}$, respectively. The overlap functions in the integrand can be classified into the following types.

The overlap in the norm kernel is given by
\begin{eqnarray}
\label{eq:overlap}
  && \bra{\Phi^{(N\Lambda)}_{n}(\mathbf{q})} \hat R(\Omega)e^{i\phi_N\hat N}e^{i\phi_Z\hat Z}\left(\frac{1+\pi\hat P}{2}\right) \ket{\Phi^{(N\Lambda)}_{n'}(\mathbf{q}')}\nonumber\\
    &=& \frac{1}{2}\left[\bra{\Phi^{(N\Lambda)}_{n}(\mathbf{q})} \hat R(\Omega)e^{i\phi_N\hat N}e^{i\phi_Z\hat Z} \ket{\Phi^{(N\Lambda)}_{n'}(\mathbf{q}')}\right.\nonumber\\
    &&\left.+ \pi\bra{\Phi^{(N\Lambda)}_{n}(\mathbf{q})} \hat R(\Omega)e^{i\phi_N\hat N}e^{i\phi_Z\hat Z} \hat P  \ket{\Phi^{(N\Lambda)}_{n'}(\mathbf{q}')}\right]. 
  \end{eqnarray}
 Substituting Eq.(\ref{eq:hypernuclear_config}) into the above expression, one finds
 \begin{eqnarray}
   &&\bra{\Phi^{(N\Lambda)}_{n}(\mathbf{q})} \hat R(\Omega)e^{i\phi_N\hat N}e^{i\phi_Z\hat Z} \ket{\Phi^{(N\Lambda)}_{n'}(\mathbf{q}')}\nonumber\\
    &=&\bra{\Phi^{(N)}(\mathbf{q})} \hat R(\Omega)e^{i\phi_N\hat N}e^{i\phi_Z\hat Z} \ket{\Phi^{(N)}(\mathbf{q}')} \nonumber\\
    &&\times  \bra{\varphi^{(\Lambda)}_n (\mathbf{q})}\hat R(\Omega)\ket{\varphi^{(\Lambda)}_{n'}(\mathbf{q}')}
\end{eqnarray}
and 
 \begin{eqnarray}
   &&\bra{\Phi^{(N\Lambda)}_{n}(\mathbf{q})} \hat R(\Omega)e^{i\phi_N\hat N}e^{i\phi_Z\hat Z}\hat P \ket{\Phi^{(N\Lambda)}_{n'}(\mathbf{q}')}\nonumber\\
    &=&\bra{\Phi^{(N)}(\mathbf{q})} \hat R(\Omega)e^{i\phi_N\hat N}e^{i\phi_Z\hat Z} \hat P\ket{\Phi^{(N)}(\mathbf{q}')} \nonumber\\
    &&\times  \bra{\varphi^{(\Lambda)}_n (\mathbf{q})}\hat R(\Omega)\hat P\ket{\varphi^{(\Lambda)}_{n'}(\mathbf{q}')}.
\end{eqnarray}
The overlaps of the hyperon part  are determined by
 \begin{subequations}
  \begin{eqnarray}
 \label{eq:D4Lambda}
 \bar D_{\Lambda_k\Lambda_{k'}}(\mathbf{q},\mathbf{q}';\Omega)
 &\equiv& \bra{\varphi^{(\Lambda)}_k (\mathbf{q})}\hat R(\Omega)\ket{\varphi^{(\Lambda)}_{k'}(\mathbf{q}')} \nonumber\\
&=&\sum_{\mu\mu^\prime}F^\ast_{k\mu}(\mathbf{q}) F_{k^\prime \mu^\prime}(\mathbf{q}') \delta_{n n'}\delta_{l l'}\delta_{j j'} D^j_{m_jm'_j}(\Omega)\nonumber \\
 && +\sum_{\mu\mu^\prime}G^\ast_{k\mu}(\mathbf{q}) G_{k^\prime \mu^\prime}(\mathbf{q}') \delta_{n n'}\delta_{l l'}\delta_{j j'} D^j_{m_jm'_j}(\Omega),\nonumber\\
\end{eqnarray}
and
 \begin{eqnarray}
 && \bra{\varphi^{(\Lambda)}_k (\mathbf{q})}\hat R(\Omega)\hat P\ket{\varphi^{(\Lambda)}_{k'}(\mathbf{q}')} \nonumber\\
&=&\sum_{\mu\mu^\prime}(-1)^{\ell}F^\ast_{k\mu}(\mathbf{q}) F_{k^\prime \mu^\prime}(\mathbf{q}') \delta_{n n'}\delta_{l l'}\delta_{j j'} D^j_{m_jm'_j}(\Omega)\nonumber \\
 && +\sum_{\mu\mu^\prime}(-1)^{\ell}G^\ast_{k\mu}(\mathbf{q}) G_{k^\prime \mu^\prime}(\mathbf{q}') \delta_{n n'}\delta_{l l'}\delta_{j j'} D^j_{m_jm'_j}(\Omega),
\end{eqnarray}
 \end{subequations}
where $F_{k\mu}$ and $G_{k \mu}$ are the expansion coefficients of the large and small components of the Dirac spinor for the $k$-th  hyperon single particle state on a set of spherical harmonic oscillator basis functions $\ket{\mu}=\ket{nljm_j}$ within 10 major shells. The $D^j_{m_jm'_j}(\Omega)=\bra{jm_j}\hat R(\Omega)\ket{jm'_j}$ is the Wigner D-function. See, for instance, Ref.~\cite{Xue:2015} for details.

 For a general one-body operator $\hat O^{(1B)}$, the corresponding overlap can be obtained by rewriting the operator in the second quantization form,
\begin{eqnarray}
   \hat O^{(1B)}= \sum_{NN'} O_{NN'}c^\dagger_N c_{N'} + \sum_{\Lambda\Lambda'}O_{\Lambda\Lambda'}c^\dagger_\Lambda c_{\Lambda'}, 
\end{eqnarray}
where the first term acts only on the wave function of the nucleons, while the second term acts on the wave function of the hyperon part. The corresponding overlap function becomes
 \begin{eqnarray}
   &&\bra{\Phi^{(N\Lambda)}_{n}(\mathbf{q})} \hat O^{(1B)} \hat R(\Omega)e^{i\phi_N\hat N}e^{i\phi_Z\hat Z} \ket{\Phi^{(N\Lambda)}_{n'}(\mathbf{q}')}\nonumber\\
    &=&\sum_{NN'} O_{NN'}\bra{\Phi^{(N)}(\mathbf{q})}c^\dagger_N c_{N'} \hat R(\Omega)e^{i\phi_N\hat N}e^{i\phi_Z\hat Z} \ket{\Phi^{(N)}(\mathbf{q}')} \nonumber\\
    &&\times  \bra{\varphi^{(\Lambda)}_n (\mathbf{q})}\hat R(\Omega)\ket{\varphi^{(\Lambda)}_{n'}(\mathbf{q}')}\nonumber\\
    &&+ \sum_{\Lambda\Lambda'}O_{\Lambda\Lambda'}\bra{\Phi^{(N)}(\mathbf{q})}  \hat R(\Omega)e^{i\phi_N\hat N}e^{i\phi_Z\hat Z} \ket{\Phi^{(N)}(\mathbf{q}')} \nonumber\\
    &&\times  \bra{\varphi^{(\Lambda)}_n (\mathbf{q})}c^\dagger_\Lambda c_{\Lambda'}\hat R(\Omega)\ket{\varphi^{(\Lambda)}_{n'}(\mathbf{q}')}.
\end{eqnarray}
 \tbd{In the spherical harmonic oscillator basis $\ket{\mu}$, the expressions for the matrix elements of the mixed densities of neutrons and protons can be found in Ref.~\cite{Yao:2009}. For the single hyperon, the matrix element of density is given by 
\begin{eqnarray}
   && \rho_{\Lambda_{\mu_2}, \Lambda_{\mu_1}}(\Omega; \mathbf{q}k; \mathbf{q}'k') \nonumber\\
   &=& \frac{\bra{\varphi^{(\Lambda)}_{k} (\mathbf{q})}c^\dagger_{\mu_1} c_{\mu_2}\hat R(\Omega)\ket{\varphi^{(\Lambda)}_{k'}(\mathbf{q}')}}
   {\bra{\varphi^{(\Lambda)}_{k} (\mathbf{q})} \hat R(\Omega)\ket{\varphi^{(\Lambda)}_{k'}(\mathbf{q}')}}\nonumber\\
   &=& \bra{\mu_2} \hat R(\Omega) \ket{\varphi^{(\Lambda)}_{k'}(\mathbf{q}')}
    \bar D^{-1}_{\Lambda_k\Lambda_{k'}}(\mathbf{q},\mathbf{q}';\Omega)
    \bra{\varphi^{(\Lambda)}_{k} (\mathbf{q})} \mu_1\rangle\nonumber\\
   &=& \bar D^{-1}_{\Lambda_k\Lambda_{k'}}(\mathbf{q},\mathbf{q}';\Omega)\Bigg(F^\ast_{k\mu_1} F_{k' \mu_2} \delta_{n_1 n_2}\delta_{l_1 l_2}\delta_{j_1 j_2} D^{j_1}_{m_{j_1}m_{j_2}}(\Omega)\nonumber\\
   &&+G^\ast_{k\mu_1} G_{k' \mu_2} \delta_{n_1 n_2}\delta_{l_1 l_2}\delta_{j_1 j_2} D^{j_1}_{m_{j_1}m_{j_2}}(\Omega)\Bigg),
\end{eqnarray}
where the $\bar D_{\Lambda_k\Lambda_{k_2}}(\mathbf{q},\mathbf{q}';\Omega)$ has been defined in (\ref{eq:D4Lambda}). The mixed density of the $\Lambda$ hyperon in coordinate space is thus given
by
\begin{eqnarray}
 &&\rho_\Lambda(\mathbf{r},\Omega;\mathbf{q}k;\mathbf{q}'k')\nonumber\\
 &=&\sum_{\Lambda_{\mu_2}\Lambda_{\mu_1}}
 \phi^\ast_{\mu_2}(\mathbf{r}) \rho_{\Lambda_{\mu_2}, \Lambda_{\mu_1}}(\Omega; \mathbf{q}k; \mathbf{q}'k')\phi_{\mu_1}(\mathbf{r}),
\end{eqnarray}
with $\phi_{\mu}(\mathbf{r})=\bra{\mathbf{r}}\mu\rangle$ being the wave function of the spherical harmonic oscillator.
}

 The overlap of a two-body operator can be evaluated similarly with the help of the generalized Wick theorem~\cite{Balian:1969}.  
More details on the calculation of the overlap functions can be found, for instance, in Refs.~\cite{Yao:2009,Yao:2022PPNP}.

The solution of the HWG equation (\ref{HWE}) provides the energy $E^{JK\pi}_{\alpha}$
and the weight function $f^{JK\pi}_{\alpha}(\mathbf{q}',n')$
for each hypernuclear state.  
 With the hypernuclear wave function (\ref{GCM:wf}), one can calculate the electric multipole ($\lambda=2, 3$) transition strength as follows,
 \begin{eqnarray}
 B(E\lambda; J^{\pi_i}_{\alpha_i} \rightarrow  J^{\pi_f}_{\alpha_f})
 \equiv\frac{1}{2J_i+1} \Bigg|M(E\lambda; J^{\pi_i}_{\alpha_i} \rightarrow  J^{\pi_f}_{\alpha_f})\Bigg|^2.
 \end{eqnarray}
 The reduced transition matrix element $M(E\lambda)$ is given by
  \begin{eqnarray}
 &&M(E\lambda; J^{\pi_i}_{\alpha_i} \rightarrow  J^{\pi_f}_{\alpha_f})\nonumber\\
 &=& \sum_{\mathbf{q_i}\mathbf{q_f};n_in_f} f^{J_fK_f\pi_f\ast}_{\alpha}(\mathbf{q}_f,n_f)f^{J_iK_i\pi_i}_{\alpha}(\mathbf{q}_i,n_i)  \nonumber\\
 &&\times \bra{NZ J_fK_f \pi_f; \mathbf{q}_f,n_f}\vert \hat Q^{(e)}_{\lambda}\vert\ket{NZ J_iK_i \pi_i; \mathbf{q}_i,n_i}, 
 \end{eqnarray}
 where the configuration-dependent reduced matrix element reads
 \begin{eqnarray}
 \label{eq:RME}
&& \bra{NZ J_fK_f \pi_f; \mathbf{q}_f,n_f}\vert \hat Q_{\lambda}\vert\ket{NZ J_iK_i \pi_i; \mathbf{q}_i,n_i}\nonumber\\
&=&\delta_{\pi_{f} \pi_{i},(-1)^{\lambda}}  \hat{J}^2_f (-1)^{J_{f}-K_{f}} \sum_{M^\prime M^{\prime \prime}}
\left(\begin{array}{c c c}
J_{f} & \lambda & J_{i} \\
-K_{f} & M^{\prime} & M^{\prime \prime}
\end{array}\right)  \nonumber\\
&&\times\bra{\Phi^{(N\Lambda)}_{n_f}(\mathbf{q}_f)}\hat{Q}^{(e)}_{\lambda M^{\prime}} \hat{P}_{M^{\prime\prime} K_{i}}^{J_{i}}\hat{P}^{Z}\hat{P}^{N} \hat{P}^{\pi}\ket{\Phi^{(N\Lambda)}_{n_i}(\mathbf{q}_i)}
 \end{eqnarray}
 In this work, only the hypernuclear states with  $K_i=K_f=K=\Omega_\Lambda=1/2$ are considered. The configuration-dependent reduced matrix element in this case is simplified as follows.
 \begin{eqnarray}
&& \bra{NZ J_f K \pi_f; \mathbf{q}_f,n_f}\vert \hat Q_{\lambda}\vert\ket{NZ J_iK  \pi_i; \mathbf{q}_i,n_i}\nonumber\\
&=& (-1)^{J_{f}-K} \delta_{\pi_{f} \pi_{i},(-1)^{\lambda}}  
\frac{\hat{J}^2_i \hat{J}^2_f}
{2} \sum_{\mu M}
\left(\begin{array}{c c c}
J_{f} & \lambda & J_{i} \\
-K & \mu & M
\end{array}\right)\nonumber\\
&&\times\int^{\pi}_0 sin\theta d\theta
d^{J_i}_{MK}(\theta)\bra{\Phi^{(N\Lambda)}_{n_f}(\mathbf{q}_f)}\hat{Q}^{(e)}_{\lambda \mu} e^{i \hat{J_y}\theta}
\hat{P}^{Z}\hat{P}^{N}\hat{P}^{\pi_i}\ket{\Phi^{(N\Lambda)}_{n_i}(\mathbf{q}_i)},\nonumber\\
 \end{eqnarray}
 where $\hat J=\sqrt{2J+1}$. In the electric multipole operator $\hat{Q}^{(e)}_{\lambda \mu}$, the bare value of the proton charge is used.

 \section{Results and discussion}%
 \label{Sec.results}

\begin{figure}[]
\centering 
\includegraphics[width=\columnwidth]{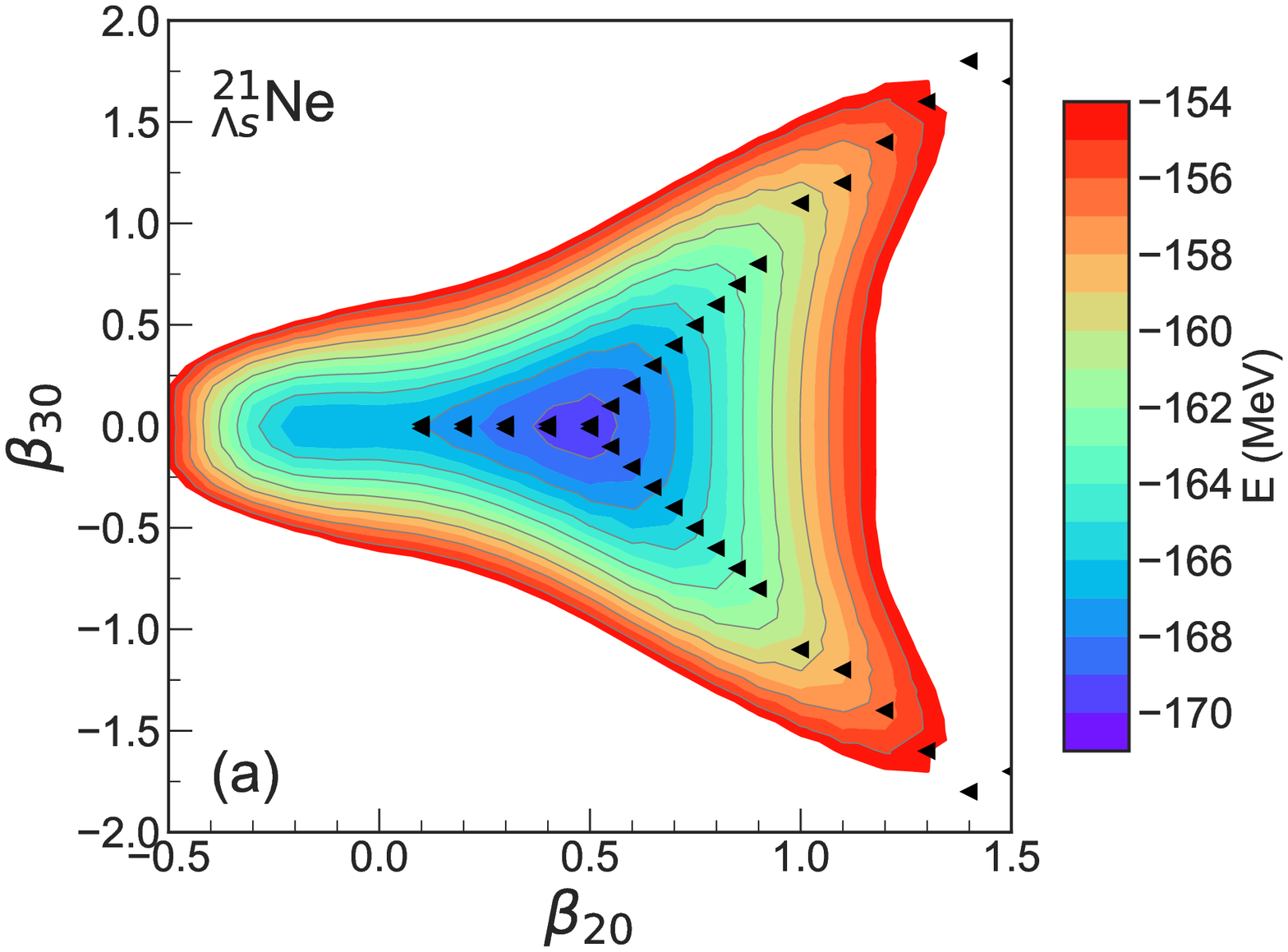}  
\includegraphics[width=\columnwidth]{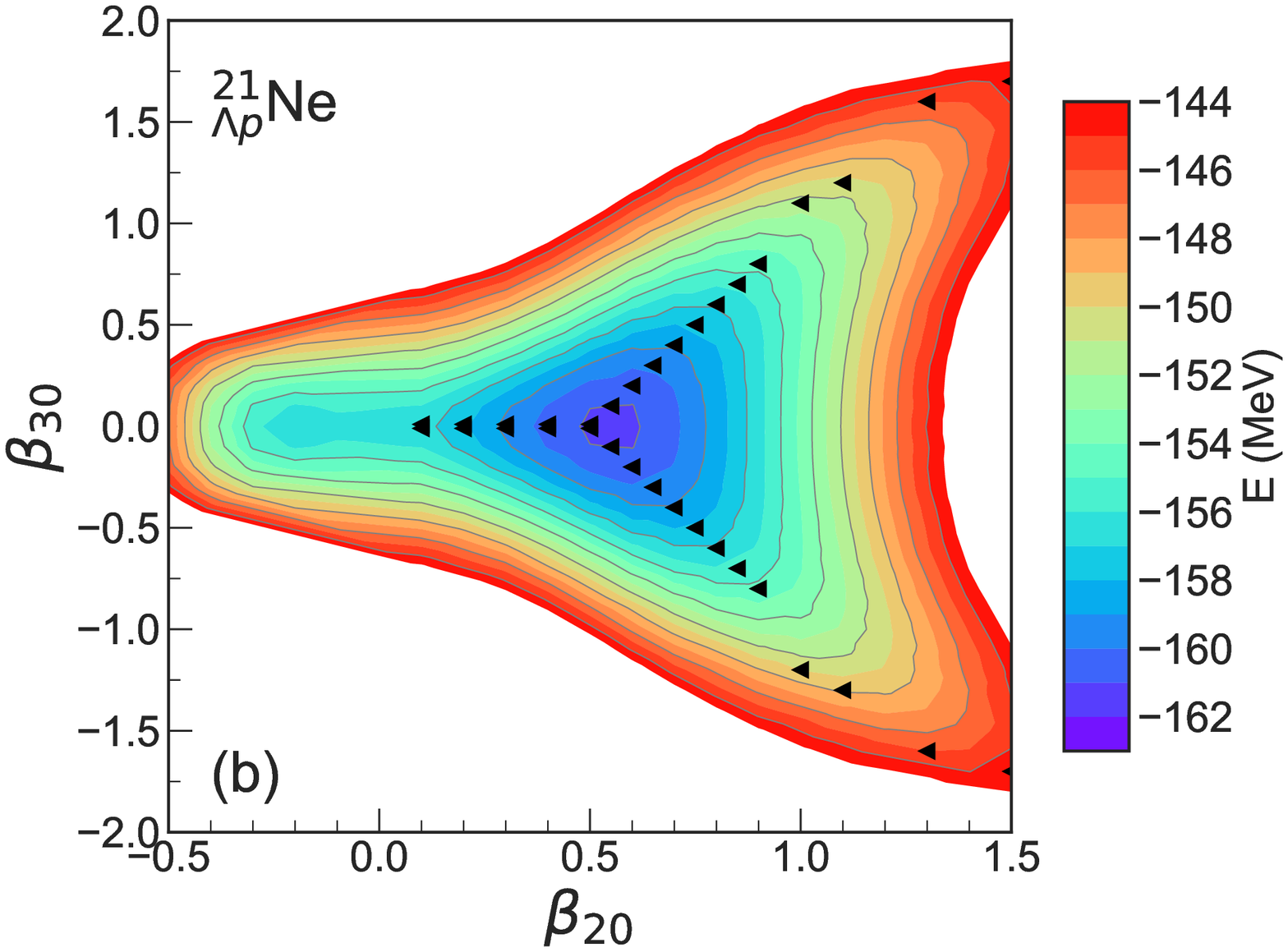}    
\caption{(Color online) The potential energy surfaces of $^{21}_{\Lambda_s}$Ne and $^{21}_{\Lambda_p}$Ne in the plane of quadrupole-octupole deformation parameters ($\beta_{20}, \beta_{30}$), where the $\Lambda$ hyperon occupies the first (a) and second (b) lowest energy states, respectively. The black triangles indicate the configurations that are employed in the HyperGCM calculations. See text for details.}
\label{figs:PES} 
\end{figure}

\begin{figure}[] 
\includegraphics[width=\columnwidth]{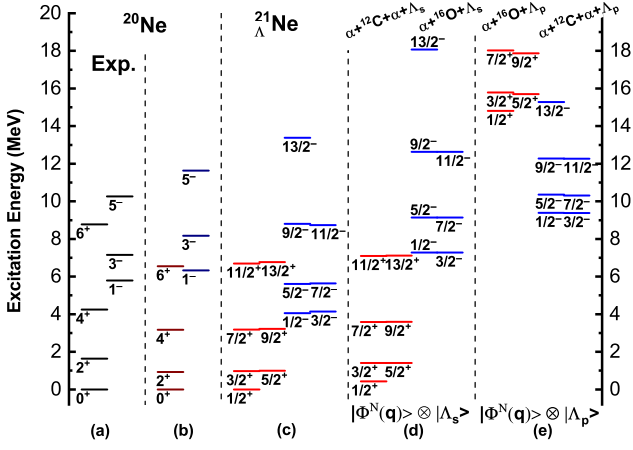}    
\caption{ The low-lying energy spectra of $^{20}$Ne (a,b) and $^{21}_{ \Lambda}$Ne (c,d,e) from the HyperGCM calculation. In the panel (d) and (e), only the configurations  $\ket{\Phi^{N}}\otimes \ket{\Lambda_{s/p}}$ with $\Lambda$ occupying either the first ($\Lambda_s$) or the second ($\Lambda_p$) lowest-energy state ($\Omega_\Lambda=1/2$) are included, respectively. In panel (c), the configurations in both (d) and (e) are allowed to be mixed.  }
\label{figs:spectra} 
\end{figure}

\begin{figure}[]
  \includegraphics[width=\linewidth]{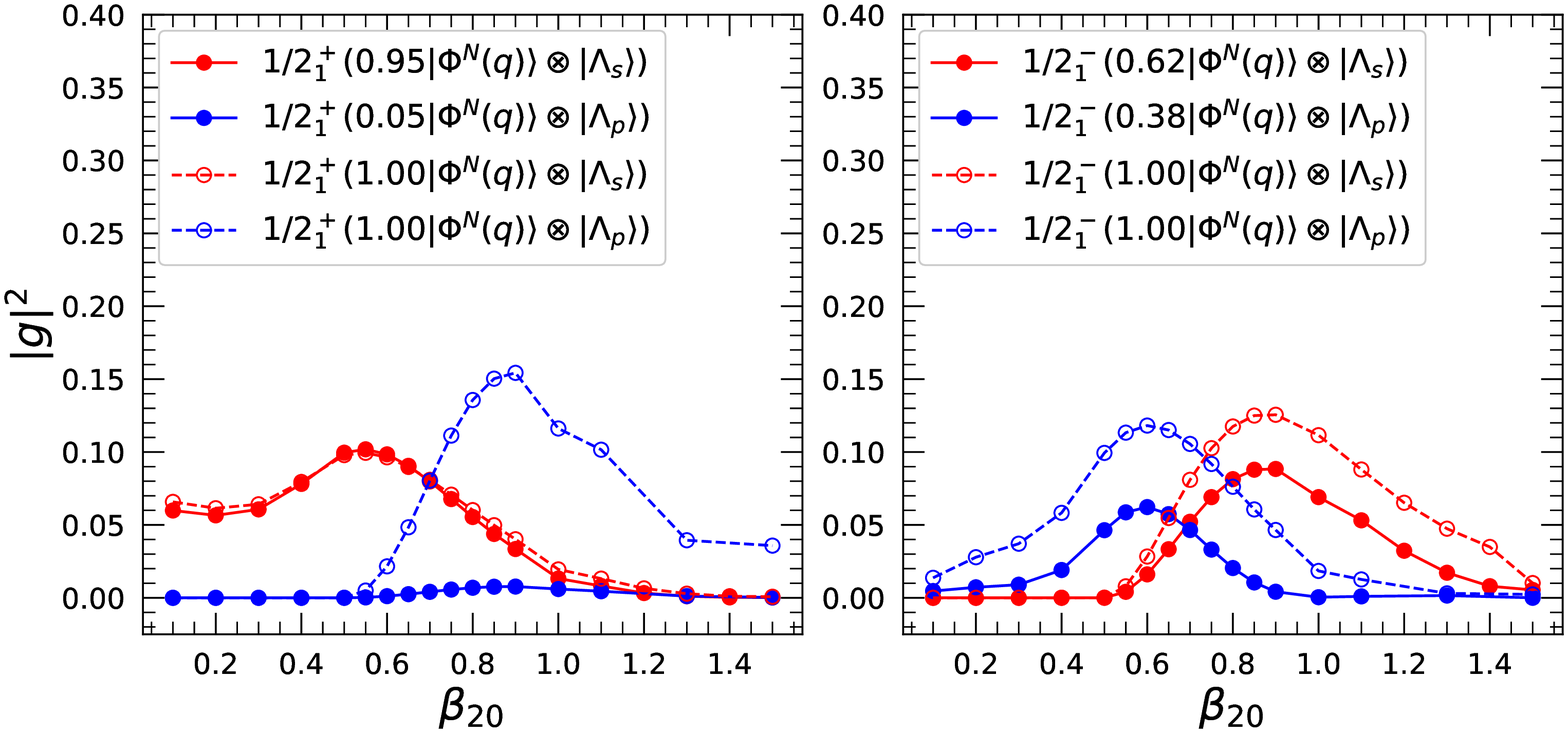}
  \includegraphics[width=\linewidth]{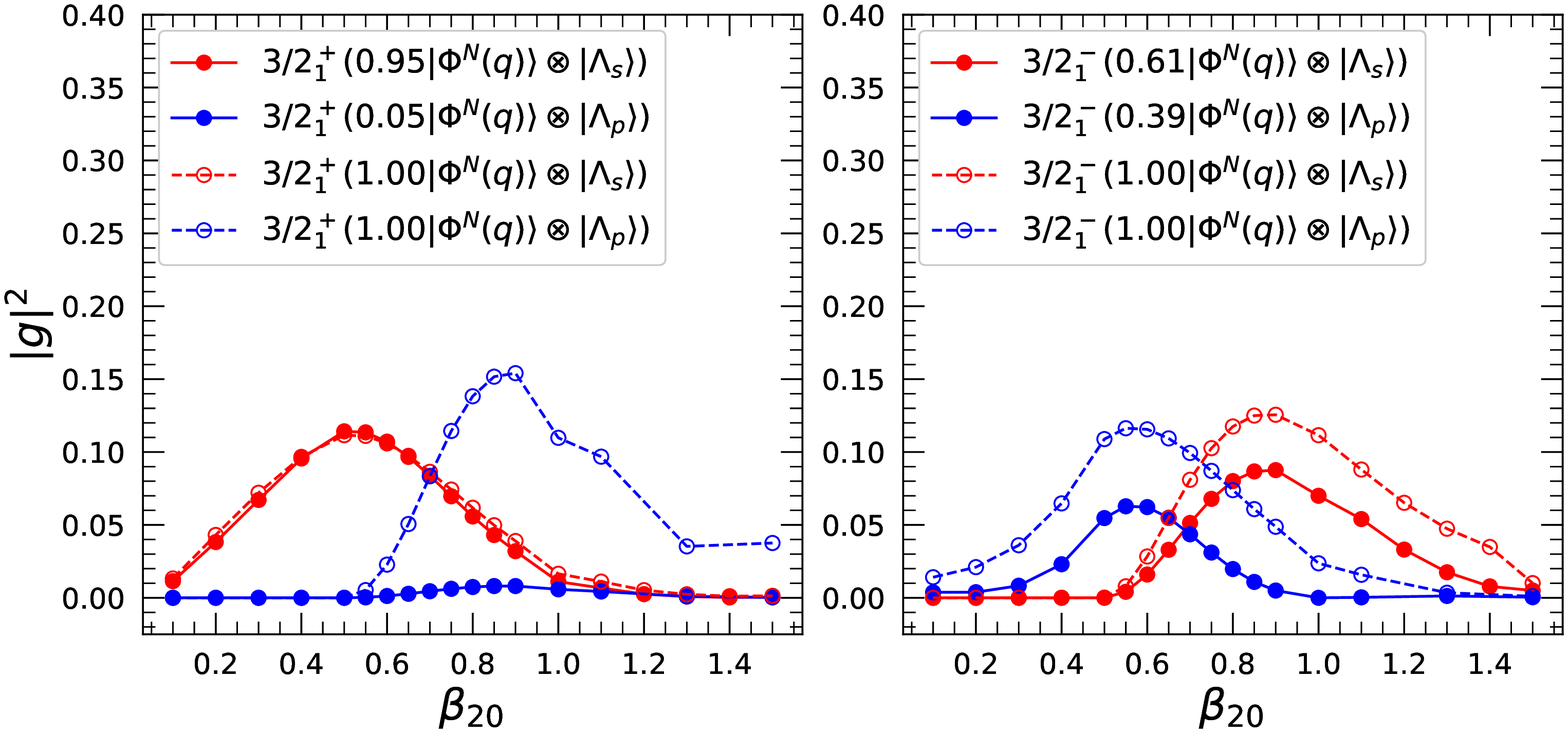} 
  \includegraphics[width=\linewidth]{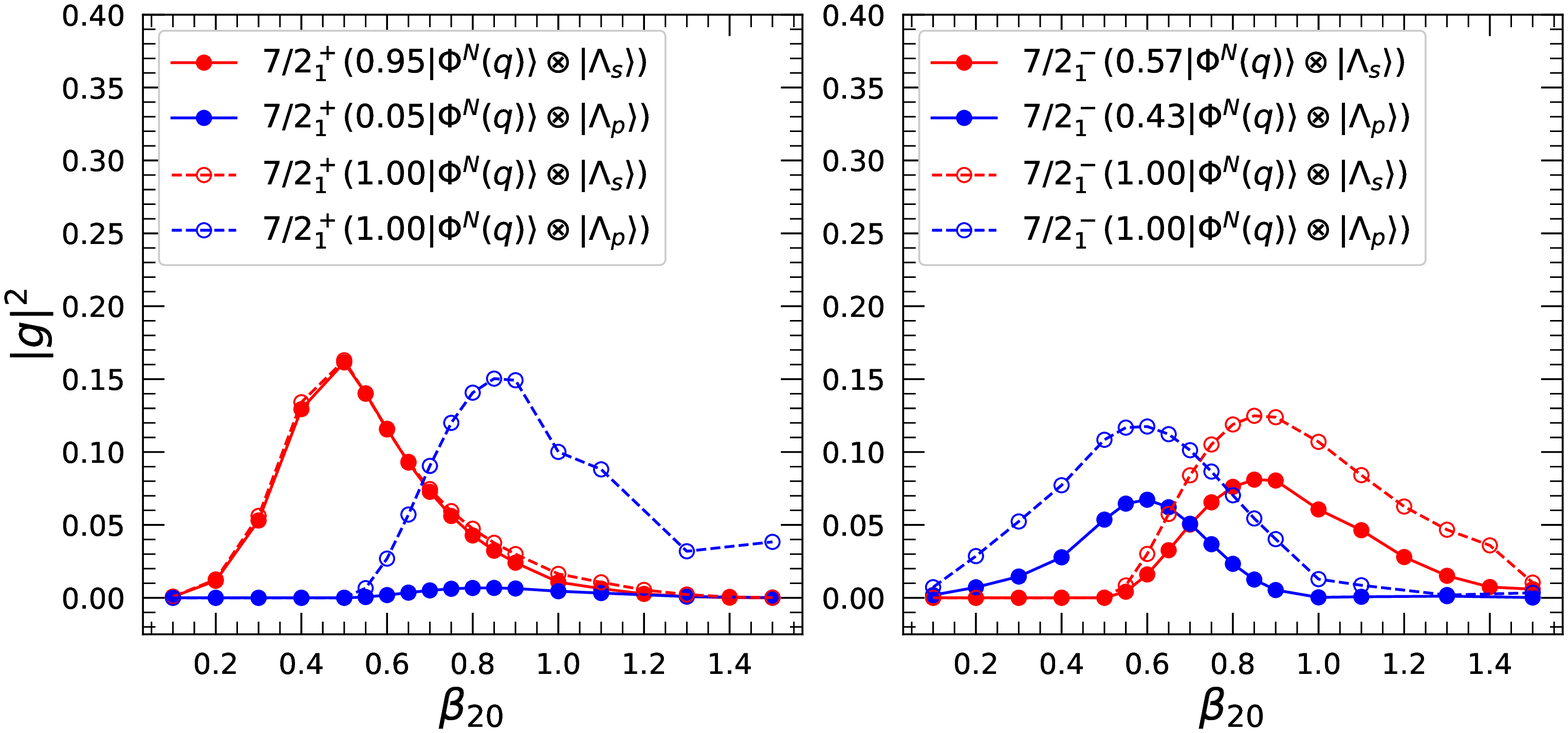}
  \includegraphics[width=\linewidth]{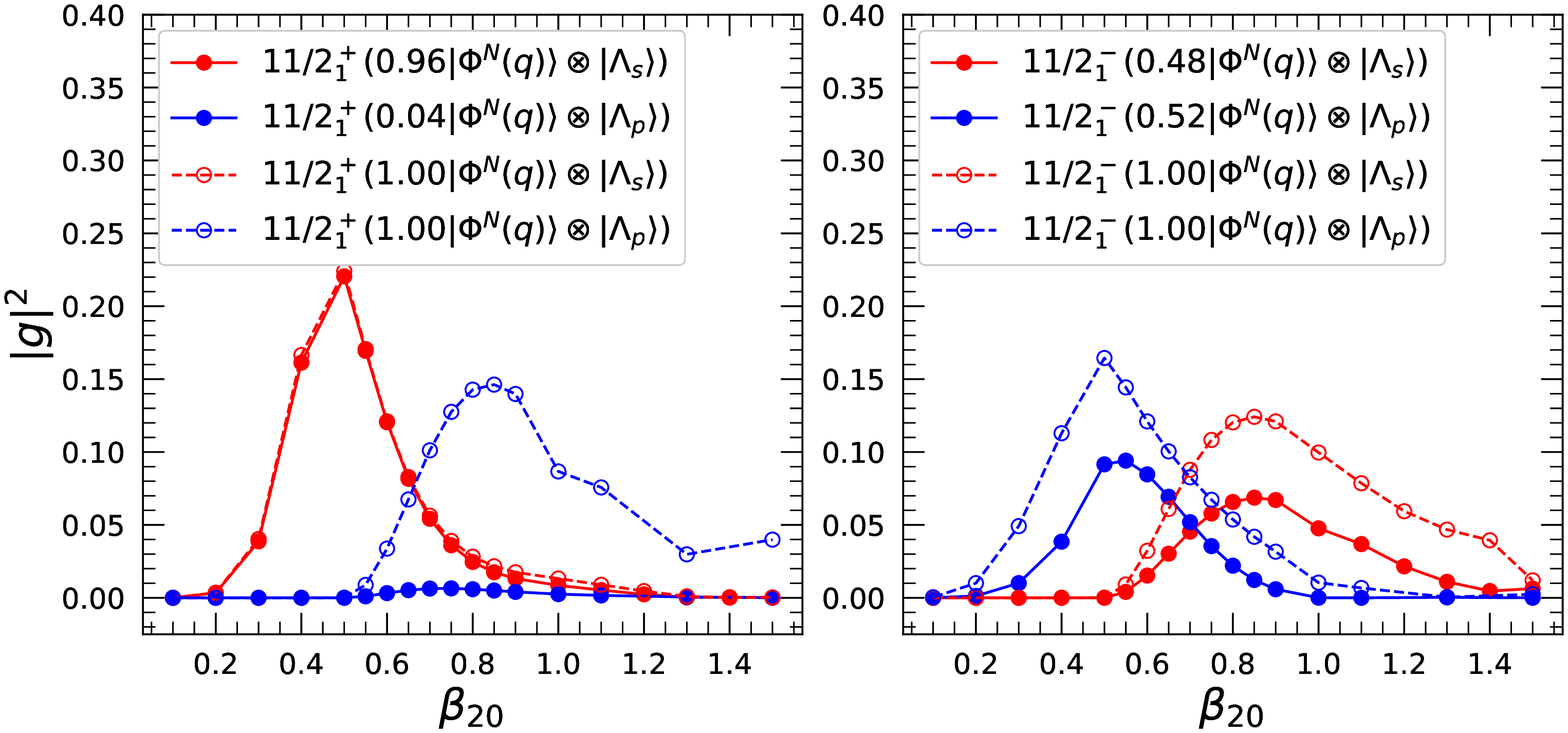}
\caption{(Color online) The distribution of collective wave functions $|g^{JK\pi}_{\alpha}(\mathbf{q},n)|^2$ for the low-lying states in Fig.~\ref{figs:spectra} from HyperGCM calculations with the mixing of different configurations. The results for the states in Fig.~\ref{figs:spectra}(d) and (e) are indicated with open symbols. The weight of each configuration in the hypernuclear state of interest is also provided.}
\label{figs:WFS}
\end{figure}

\begin{figure}[]
  \includegraphics[height=3.8cm]{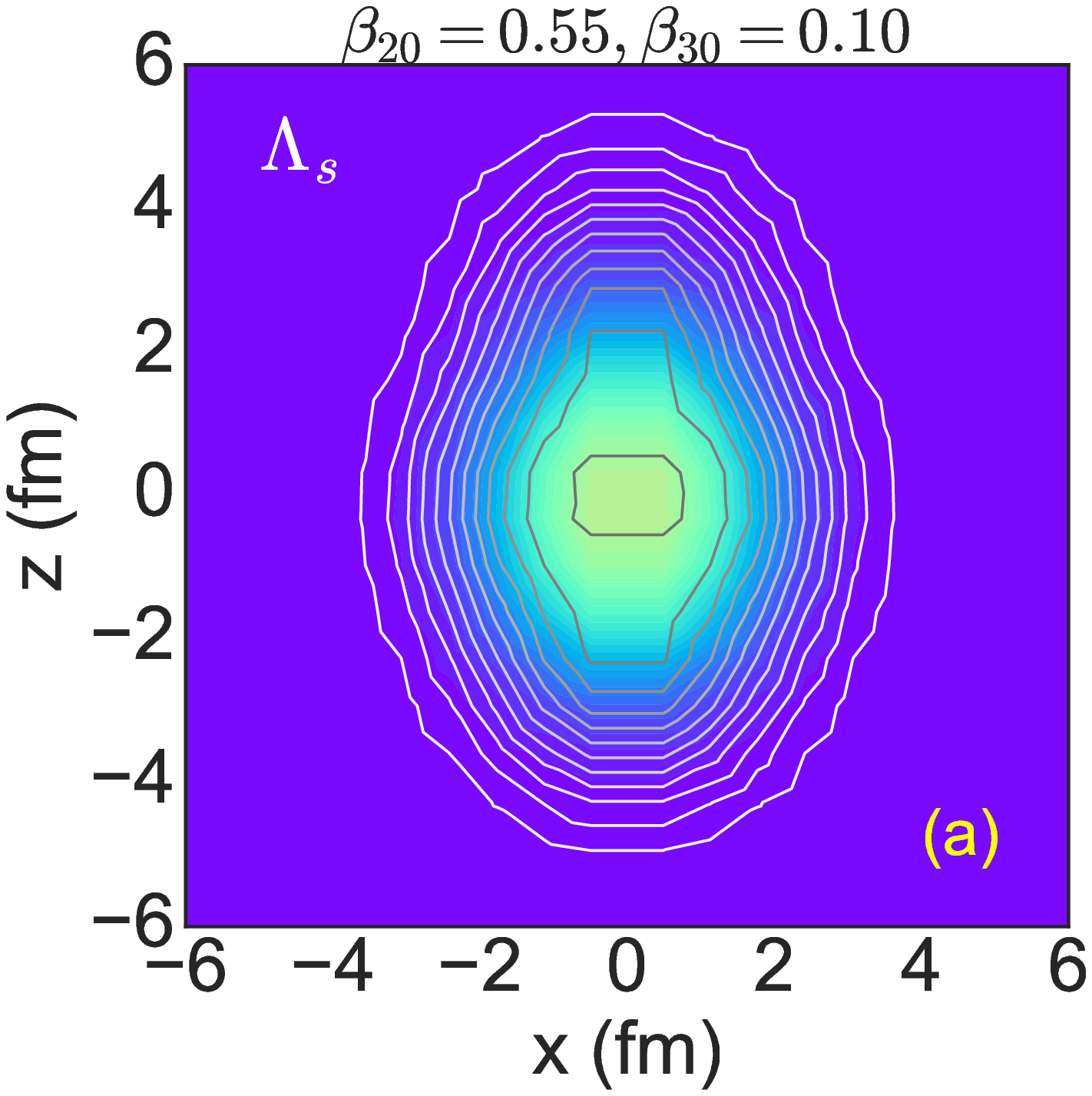}
  \includegraphics[height=3.8cm]{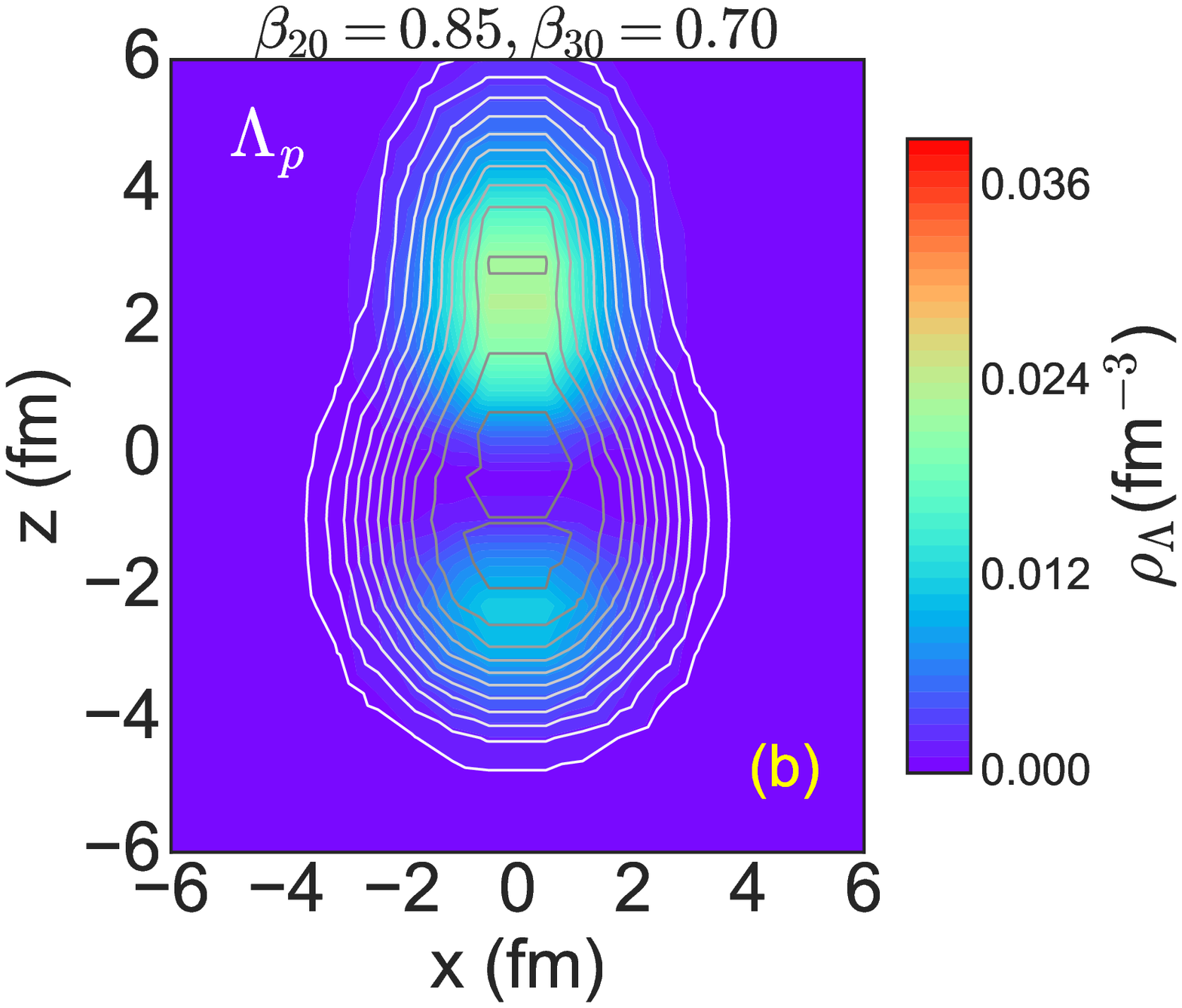}  
\caption{(Color online) The contour plots of the nucleon density  and the density profile of the $\Lambda$ in the $(x, z)$ plane at $y = 0$ fm (a) for the  configuration $\ket{\Phi^{N}}\otimes \ket{\Lambda_{s}}$ with $\beta_{20}=0.55, \beta_{30}=0.10$ [which dominates the positive-parity states in Fig.~\ref{figs:spectra}(c)], and (b) for the configuration $\ket{\Phi^{N}}\otimes \ket{\Lambda_{p}}$ with $\beta_{20}=0.85, \beta_{30}=0.70$. The difference between two neighboring lines is $0.015$ fm$^{-3}$. }
\label{figs:density_positive}
\end{figure}

\begin{figure}[] 
  \includegraphics[height=3.8cm]{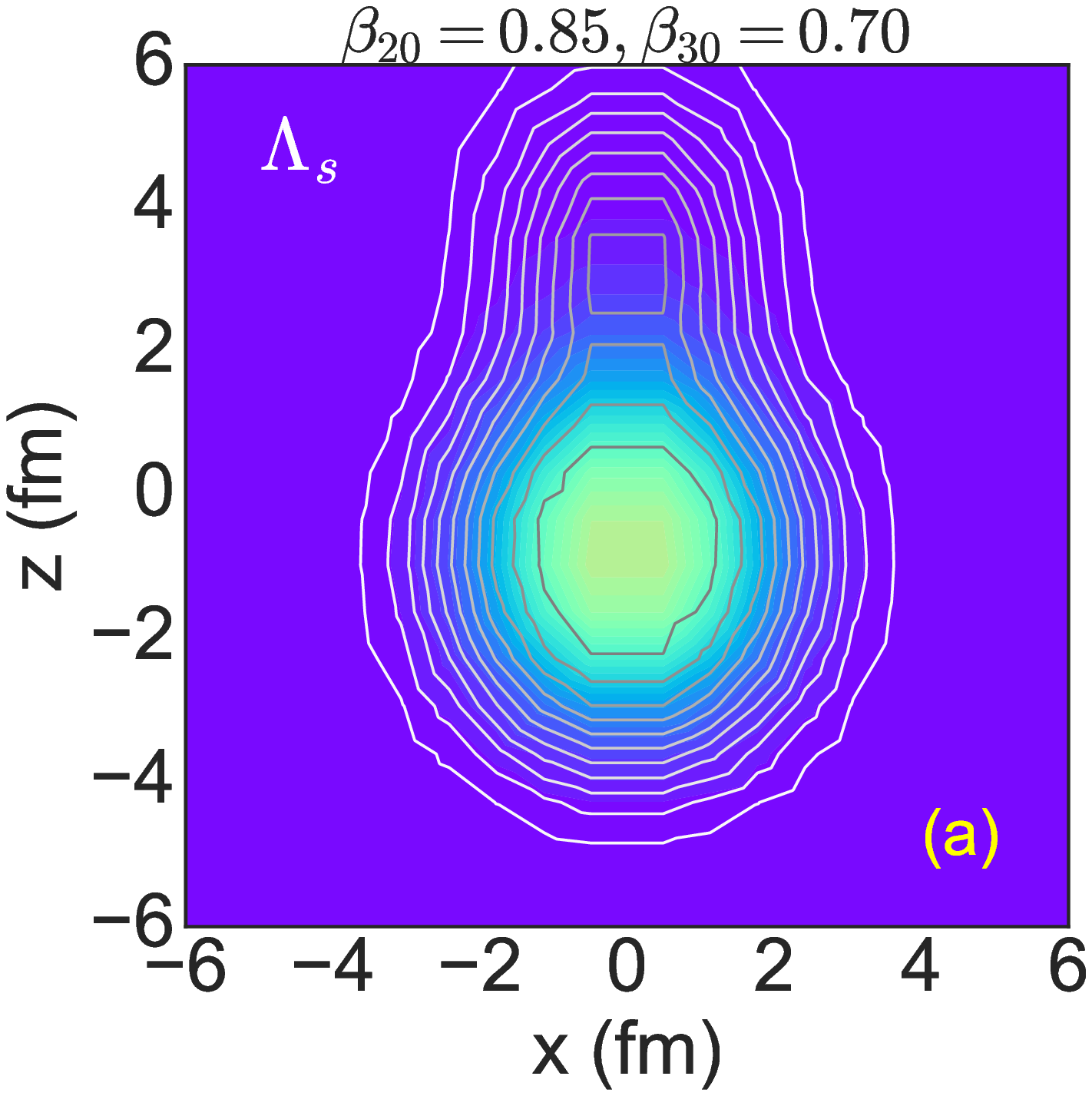}
  \includegraphics[height=3.8cm]{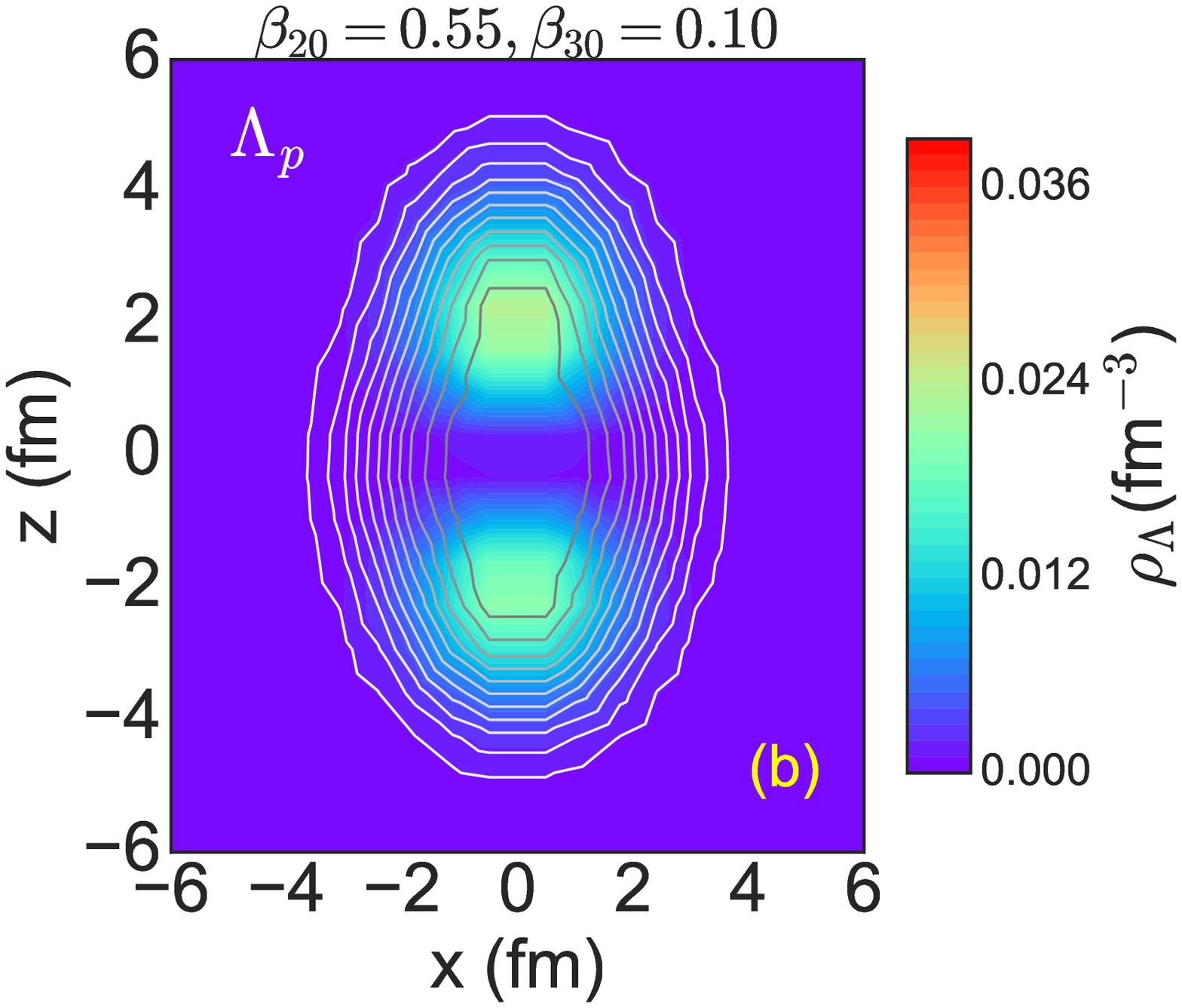}
\caption{(Color online) Same as Fig.~\ref{figs:density_positive}, but (a) for the configuration $\ket{\Phi^{N}}\otimes \ket{\Lambda_{s}}$ with $\beta_{20}=0.85, \beta_{30}=0.70$ and (b) for the configuration $\ket{\Phi^{N}}\otimes \ket{\Lambda_{p}}$  with $\beta_{20}=0.55, \beta_{30}=0.10$. These two configurations are strongly mixed in the negative-parity states in Fig.~\ref{figs:spectra}(c).}
\label{figs:density_negative}
\end{figure}

Figure~\ref{figs:PES}  displays the potential energy surfaces (PESs)  of $^{21}_{ \Lambda}$Ne, where the $\Lambda$ hyperon occupies the first and second lowest-energy states with $K_\Lambda=1/2$, respectively. These PESs are essentially the same as those in Ref.~\cite{Xia:2019}, except that in the present work the quadrupole deformation $\beta_{20}$ and octupole deformation $\beta_{30}$ are extended up to $\beta_{20}=1.5$ and $\beta_{30}=2.0$, respectively. The contour lines of the PESs form triangles approximately with the center located around $\beta_{20}=0.55, \beta_{30}=0.0$, which corresponds to the global energy minimum.

The low-lying states of $^{21}_{ \Lambda}$Ne are obtained with the HyperGCM, where the quadrupole-octupole deformation parameters $(\beta_{20}, \beta_{30})$ are chosen as generator coordinates. Considering the observation in Ref.~\cite{Zhou:2016PLB} that  the low-lying energy spectrum predicted by the GCM with the mixing of  all the configurations on the entire $(\beta_{20}, \beta_{30})$ plane can be reasonably reproduced by mixing the configurations only located along the ``valley", for the sake of simplicity, only the configurations indicated with black triangles in Fig.~\ref{figs:PES} are included in the present HyperGCM calculation. \tbd{Considering the fact that the configurations with $\beta_{30}<0$ are automatically included in the HyperGCM calculation with the parity-projection operator, only the configurations with $\beta_{30}\ge0$ are needed in the practical calculation.}
The predicted low-lying states for $^{21}_{ \Lambda}$Ne are shown in Fig.~\ref{figs:spectra}(c). For comparison, the data and GCM predicted energy spectra of $^{20}$Ne are shown in Fig.~\ref{figs:spectra}(a) and (b), respectively. Those of $^{21}_{ \Lambda}$Ne from the HyperGCM calculations with the mixing of the configurations of only $^{21}_{ \Lambda_s}$Ne or $^{21}_{ \Lambda_p}$Ne are shown in Fig.~\ref{figs:spectra}(d), and (e), respectively. It is worth pointing out that in our previous work~\cite{Mei:2016PRC} with the mixing of only axially deformed configurations, the predicted hypernuclear low-lying states with $K=1/2$ should be compared to the positive-parity states in Fig.~\ref{figs:spectra}(d) and the negative-parity states in Fig.~\ref{figs:spectra}(e). Here, octupole-deformed configurations are included additionally. As a result, we also obtain negative-parity states from the mixing of configurations $\ket{\Phi^{N}}\otimes \ket{\Lambda_s}$, and positive-parity states from the mixing of configurations $\ket{\Phi^{N}}\otimes \ket{\Lambda_p}$. Since parity is violated in the configurations of both  $^{21}_{ \Lambda_s}$Ne and $^{21}_{ \Lambda_p}$Ne, these two types of configurations can also mix, leading to the spectrum in Fig.~\ref{figs:spectra}(c). Because the mixing introduces additional correlations, each state in Fig.~\ref{figs:spectra}(c) is lower than that in both Fig.~\ref{figs:spectra}(d) and Fig.~\ref{figs:spectra}(e). This is particularly true for the negative-parity states. Comparing (c) and (d), it is seen that the excitation energies of the positive-parity doublets $(3/2^+, 5/2^+)$ are shifted down by about 0.4 MeV, while those of the negative-parity doublets $(1/2^-, 3/2^-)$ are reduced by about 3.2 MeV. We note that the energy splitting between $(1/2^-, 3/2^-)$ is 88 keV, which cannot be interpreted as the spin-orbit splitting of the hyperon $p$ state as their wave functions are far more complicated than the picture of a spherical nuclear core coupled to $p$-orbital $\Lambda$~\cite{Xia:2017}. The size of energy splitting is much smaller than that found in the HyperGCM calculation~\cite{Mei:2016PRC} with the mixing of only axially deformed configurations $\ket{\Phi^N}\otimes\ket{\Lambda_p}$. Besides, we note that the energy splitting between the positive-parity doublets $(3/2^+, 5/2^+)$ and $(7/2^+, 9/2^+)$ is 18 keV and 38 keV, respectively.

\begin{table*}[tb]
\centering
    \tabcolsep=5pt 
    \caption{The electric quadrupole transition strengths $B(E2)$  for the low-lying states in $^{20}$Ne and $^{21}_\Lambda$Ne from the (Hyper)GCM calculation starting from the relativistic point-coupling PC-F1 plus PCY-S2 interactions, in comparison with those by the HyperAMD model~\cite{Isaka:2011}.  The results from the HyperGCM calculations with the mixing of one of the two types of configurations $\ket{\Phi^{N}}\otimes \ket{\Lambda_{s}}$ and $\ket{\Phi^{N}}\otimes \ket{\Lambda_{p}}$ are also given for comparison, and they are labeled with HyperGCM($\Lambda_s$) and HyperGCM($\Lambda_p$), respectively. The data for $^{20}$Ne is taken from Ref.~\cite{NNDC}. }
    \begin{tabular}{cccccccccccc}
      \hline \hline
        \multicolumn{4}{c}{$^{20}$Ne}  &     \multicolumn{5}{c}{$^{21}_\Lambda$Ne} \\
         \cline{1-4}   \cline{6-10} \\
        & \multicolumn{4}{c}{$B(E2;I^\pi_i\to I^\pi_f)$ ($e^2$fm$^4$)}    &    \multicolumn{6}{c}{$B(E2;J^\pi_i\to J^\pi_f)$ ($e^2$fm$^4$)} & \\
         \cline{2-4}  \cline{6-10} \\
        $I^\pi_i\to I^\pi_f$ & Exp. & GCM  & AMD~\cite{Isaka:2011}  &  & $J^\pi_i\to J^\pi_f$  & HyperGCM  & HyperGCM($\Lambda_s$) & HyperGCM ($\Lambda_p$)  & HyperAMD~\cite{Isaka:2011}   \\
        \hline
        $2^+_1\to 0^+_1$ &63(5)& 72.3 & 72.2 &    & $3/2^+_1\to 1/2^+_1$  &58.7&59.0 &108.7 & 63.7  \\
                         && &  & & $5/2^+_1\to 1/2^+_1$  &58.7 &59.0&107.1 & 63.9    \\
                         
        $4^+_1\to 2^+_1$ &71(6) & 94.1 & 86.9&    & $7/2^+_1\to 3/2^+_1$  & 73.0&73.3&139.9 & 64.3   \\
                         & && & &   $9/2^+_1\to 5/2^+_1$  & 75.8 &76.4&148.9 &  75.7   \\
        $6^+_1\to 4^+_1$ & 64(10) & 79.7 & 55.1&  & $11/2^+_1\to 7/2^+_1$  &71.4 &71.9 &159.1 & 40.3   \\
                         & && &  & $13/2^+_1\to 9/2^+_1$  &70.4 &71.1 &159.6 &  48.0  \\
       \hline 
               $3^-_1\to 1^-_1$ & 164(26) &150.0   & 221.2  &  & $5/2^-_1\to 1/2^-_1$  &102.3 &140.2 & 65.9 &139.2 &     \\
                         && & &   & $7/2^-_1\to 3/2^-_1$  & 114.1 & 163.8 &67.6 & 178.5\\
                         
        $5^-_1\to 3^-_1$ && 172.3  & 249.3 &  & $9/2^-_1\to 5/2^-_1$  &120.3 &187.3 &68.6 &  184.2   \\ 
                         && & &&   $11/2^-_1\to 7/2^-_1$  &122.6  & 193.7 & 71.0&    189.3 \\
        $7^-_1\to 5^-_1$ &&178.1    & 240.3&  & $13/2^-_1\to 9/2^-_1$  &112.6 & 211.7 & 64.5& 166.7   \\
       \hline \hline
    \end{tabular}
    \label{tab:BE2}
\end{table*}

\begin{table*}[tb]
    \centering
    \tabcolsep=12pt 
    \caption{Same as Tab.~\ref{tab:BE2}, but for the electric octupole transition strengths $B(E3)$($e^2$fm$^6$). The absolute value of the reduced matrix element $M(E3)=|\bra{J_f}|Q_3|\ket{J_i}|$ ($e$fm$^3$) defined in (\ref{eq:RME}) is given in parentheses for comparison. Data for $^{20}$ Ne are taken from~\cite{Kibedi:2002BE3}. }
    \begin{tabular}{cccccrrl}
      \hline \hline
        \multicolumn{3}{c}{$^{20}$Ne}  & &    \multicolumn{3}{c}{$^{21}_\Lambda$Ne} \\
         \cline{1-3}   \cline{5-7} \\
        & \multicolumn{2}{c}{$B(E3;I^\pi_i\to I^\pi_f$) ($M(E3)$)}  &  & &   \multicolumn{2}{c}{$B(E3;J^\pi_i\to J^\pi_f$)  ($M(E3)$)}   \\
         \cline{2-3}   \cline{6-7} \\
        $I^\pi_i\to I^\pi_f$ & Exp. & GCM &  & $J^\pi_i\to J^\pi_f$  & HyperGCM  & HyperGCM($\Lambda_s$)  \\
        \hline
        $3^-_1\to 0^+_1$       &   321 & 257.5 (42.5)   & & 
        $5/2^-_1\to 1/2^+_1$   &  69.7 (20.5) & 205.3 (35.1)   & \\ 
                   &   &     & & 
        $7/2^-_1\to 1/2^+_1$   & 48.9 (19.8) & 160.3 (35.8)    & \\ 
        $5^-_1\to 2^+_1$       &   & 330.9 (60.3)   & & 
        $9/2^-_1\to 3/2^+_1$   & 91.5 (30.2) & 327.6 (57.2)   & \\ 
                   &   &     & & 
        $11/2^-_1\to 5/2^+_1$   & 47.4 (23.9) & 213.2 (50.6)    & \\ 
        $1^-_1\to 4^+_1$       &  & 677.8 (45.1)   & & 
        $1/2^-_1\to 7/2^+_1$   & 497.0 (31.5)  & 1379.8 (52.5) & \\ 
                   &   &     & & 
        $3/2^-_1\to 9/2^+_1$   & 240.1 (31.0)  & 660.5 (51.4) & \\ 
       \hline \hline
    \end{tabular}
    \label{tab:BE3}
\end{table*}

Decomposing nuclear wave functions helps to understand how each configuration contributes to the states of interest. Since the weight functions,  $f^{JK\pi}_{\alpha}(\mathbf{q},n)$ in Eq.(\ref{GCM:wf})  are not orthogonal to each other and their module squares cannot be interpreted as a probability, a new collective wave function $g^{JK\pi}_{\alpha}(\mathbf{q},n)$ is usually introduced as follows,
 \begin{equation} 
g^{JK\pi}_{\alpha}(\mathbf{q},n)
=\sum_{n',\mathbf{q}'}  \left[{\cal N}^{JK\pi}_{nn'}(\mathbf{q},\mathbf{q}')\right]^{1/2}f^{JK\pi}_{\alpha}(\mathbf{q}',n'),
 \end{equation} 
 which fulfills the normalization condition. 
 The modules of the collective wave functions $|g^{JK\pi}_{\alpha}(\mathbf{q},n)|^2$ from different types of calculations for the low-lying hypernuclear states in Fig.~\ref{figs:spectra}(c), (d), and (e) are shown in Fig.~\ref{figs:WFS}. One can see that the low-lying positive-parity states are dominated by the configurations
 $\ket{\Phi^{N}}\otimes \ket{\Lambda_s}$ which exhaust more than 95\% of the total wave function. In contrast, the negative-parity states are strong admixtures of the configurations  $\ket{\Phi^{N}}\otimes \ket{\Lambda_s}$ and  $\ket{\Phi^{N}}\otimes \ket{\Lambda_p}$. In the lowest negative-parity states $(1/2^-, 3/2^-)$, the ratio of the component $\ket{\Phi^{N}}\otimes \ket{\Lambda_s}$ to the component $\ket{\Phi^{N}}\otimes \ket{\Lambda_p}$ is about two. With the increase of angular momentum, the mixing becomes stronger with the ratio close to one. This strong admixture causes the evident energy shift of the negative-parity states. This is different from what has been found in the previous studies~\cite{Isaka:2011,Mei:2016PRC,Xia:2019}. Similar to the finding in $^{20}$Ne~\cite{Zhou:2016PLB}, the peak position of the $|g|^2$ for the positive-parity states of $^{21}_\Lambda$Ne is slightly decreasing with the increase of angular momentum $J$, indicating the quadrupole collectivity is weakening at the higher-spin states. 
 
  \tbd{Clustering structure in $^{20}$Ne has been comprehensively studied in nuclear EDFs~\cite{Kimura:2004PRC,Ebran:2012Nature,Ebran:2014PRC} and beyond mean-field approaches~\cite{Zhou:2013PRL,Zhou:2016PLB}. It has been found in Refs.~\cite{Kimura:2004PRC,Zhou:2016PLB} that the positive-parity states of $^{20}$Ne are dominated by the $\alpha+^{12}$C+$\alpha$ structure, while the negative-parity states are by the pear-shaped $\alpha+^{16}$O structure. The occurrence of clustering structure in nuclear core may change evidently the binding energy of hyperons in hypernuclei~\cite{Lu:2014PRC,Isaka:2015PTEP,Wu:2017PRC,Cui:2022CPC}.}  Figs.~\ref{figs:density_positive} and \ref{figs:density_negative} show the density profiles of nucleons and $\Lambda$ hyperon of the predominant configurations for the positive and negative parity states, respectively. With the inclusion of one $\Lambda$ hyperon, the positive-parity states of $^{21}_\Lambda$Ne are anticipated to be dominated by  the $\alpha+^{12}$C+$\alpha+\Lambda_s$ structure, as shown in Fig.~\ref{figs:density_positive}. In contrast,  the negative-parity states are strong admixtures of $\alpha+^{16}$O+$\Lambda_s$ structure and $\alpha+^{12}$C+$\alpha+\Lambda_p$ structure, as discussed in Fig.~\ref{figs:WFS}.

Tables~\ref{tab:BE2} and ~\ref{tab:BE3} list the transition strengths of the electric quadrupole ($E2$) and octupole ($E3$) for the low-lying states in $^{21}_\Lambda$Ne.  For comparison, the transition strengths in $^{20}$Ne from the GCM calculation based on the PC-F1 force for the $NN$ interaction, and those in $^{21}_\Lambda$Ne from the HyperAMD calculation~\cite{Isaka:2011} are also provided. It is seen from Tab.~\ref{tab:BE2}  that the $B(E2; 2^+_1\to 0^+_1)$  in $^{20}$Ne and the $B(E2)$ values for the transitions from $(3/2,5/2)^+_1\to 1/2^+_1$  in $^{21}_\Lambda$Ne by the HyperGCM and HyperAMD calculations are consistent with each other. Both results demonstrate the impurity effect of $\Lambda$ on the reduction of nuclear quadrupole collective properties. In particular, the feature that the $B(E2)$ value first increases and then decreases with angular momentum in $^{20}$Ne ($^{21}_\Lambda$Ne) is reproduced (predicted) by both methods, even though the values are quantitatively different from each other. Moreover, one can see that the $B(E2)$ values  by the HyperGCM with the mixing of $\Lambda_s$ and $\Lambda_p$ orbits are systematically and slightly smaller than those by the HyperGCM($\Lambda_s$), indicating that the mixing of the component $\ket{\Phi^{N}}\otimes \ket{\Lambda_p}$ additionally also reduces nuclear quadrupole collectivity. This reduction effect is shown to be more obvious in the negative-parity states than in the positive-parity states.  Tab.~\ref{tab:BE3} shows that this reduction effect is stronger in the $B(E3)$ values. Quantitatively, the reduced $E3$ matrix element for the transition $5/2^-\to 1/2^+$ is quenched from 35.1 $e$fm$^3$ to 20.5  $e$fm$^3$, corresponding to the quenching factor of $42\%$. 

\section{Summary}
\label{Sec.summary}
We have extended the HyperGCM framework for the low-lying hypernuclear states with the mixing of the configurations associated with both the hyperon excitation and quadrupole-octupole collective excitations based on a covariant density functional theory. The method has been applied to the low-lying states of $^{21}_\Lambda$Ne which are dominated by clustering structure.  We have found that the inclusion of additional octupole-deformed configurations leads to the negative-parity states with a strong mixing of the configurations where the $\Lambda$ hyperon occupies the first and second lowest-energy states, respectively. In other words, the low-lying negative-parity states are dominated by the admixtures of $\alpha+^{16}$O+$\Lambda_s$ structure and $\alpha+^{12}$C+$\alpha+\Lambda_p$ structure.  As a result, the excitation energies of low-lying negative-parity states become much smaller than what are expected from the previous studies~\cite{Isaka:2011,Cui:2017,Mei:2016PRC}. Besides, the admixture of these two types of configurations leads to a reduction in the values of $B(E2)$. This reduction effect is even stronger in the $E3$ transition strengths which is to be confirmed in the future hypernuclear experiments. In contrast, the positive-parity states are still dominated by the $\alpha+^{12}$C+$\alpha+\Lambda_s$ structure.  This newly-developed HyperGCM framework provides a theoretical tool of choice to study the impact of baryon-baryon interactions on hypernuclear low-lying states, especially the to-be-measured electric and magnetic dipole transitions which are expected to be sensitive to the coupling strengths in the $\Lambda N$ interaction~\cite{Yao:2008,Sang:2013}. Work in this direction is in progress.

\section*{Acknowledgments} 
 HJX is supported by Science and Technology Project of Hebei Education Department (No. ZC2021011), Scientific Research and Development Planning Project of Handan City (No. 21422901160). XYW is supported by the National Natural Science Foundation of China
under Grant No. 12005082, the Jiangxi Provincial Natural Science Foundation 20202BAB211008, Jiangxi Normal University (JXNU) Initial Research Foundation Grant to Doctor (12019504), and the Young Talents Program under JXNU (12019870). JMY is partially supported by Natural Science Foundation under Grant No. 12141501 and the Fundamental Research Funds for Central Universities, Sun Yat-sen University.
 


%

\end{document}